\documentclass[twocolumn,letterpaper]{article}

\usepackage[T1]{fontenc}
\usepackage{amsmath}
\usepackage{graphicx}
\usepackage{hyperref}
\usepackage{algorithmic}
\usepackage{booktabs}
\usepackage{multirow}
\usepackage{listings}
\usepackage{caption}
\usepackage{float}
\usepackage{xspace}
\usepackage{color}
\usepackage{textcomp}
\usepackage{xcolor}
\usepackage{multicol}

\usepackage[margin=1in]{geometry}
\newcommand{\gpto}{GPT-3.5o\xspace}
\newcommand{\gptn}{GPT-3.5n\xspace}
\newcommand{\gptt}{GPT-4\xspace}
\newcommand{\mixmoe}{Mix-8x7B\xspace}
\newcommand{\mixlarge}{Mix-8x22B\xspace}
\newcommand{\gpro}{GPro-1.0\xspace}

\newfloat{code}{htbp}{lop}
\floatname{code}{Listing}

\lstdefinelanguage{Rust}{
    keywords={let, mut, fn, struct, enum, impl, for, in, if, else, match, return, loop, continue, break, as, use, pub, mod, crate, extern, const, static, trait, type, unsafe, super, self, while, true, false},
    keywordstyle=\color{blue}\bfseries,
    ndkeywords={Option, Result, String, Vec, i32, usize},
    ndkeywordstyle=\color{darkgray}\bfseries,
    identifierstyle=\color{black},
    sensitive=false,
    comment=[l]{//},
    morecomment=[s]{/*}{*/},
    commentstyle=\color{green}\ttfamily,
    stringstyle=\color{red}\ttfamily,
    morestring=[b]",
}

\title{\bf Enabling New HDLs with Agents}
\author{
  \begin{tabular}{ccc}
    Mark Zakharov,&Farzaneh Rabiei Kashanaki,&Jose Renau
  \end{tabular}
    \\
    Department of Computer Science and Engineering \\
    University of California, Santa Cruz \\
    \texttt{\{mzakharo, frabieik, renau\}@ucsc.edu}
}

\begin{document}

\maketitle

\begin{abstract}

Large Language Models (LLMs) based agents are transforming the programming language
landscape by facilitating learning for beginners, enabling code generation, and
optimizing documentation workflows. Hardware Description Languages (HDLs),
with their smaller user community, stand to benefit significantly from the
application of LLMs as tools for learning new HDLs. This paper investigates the
challenges and solutions of enabling LLMs for HDLs, particularly for
HDLs that LLMs have not been previously trained on.

This work introduces HDLAgent, an AI agent optimized for LLMs with limited
knowledge of various HDLs. It significantly enhances off-the-shelf LLMs. For
example, PyRTL’s success rate improves from zero to 35\% with Mixtral 8x7B, and
Chisel’s success rate increases from zero to 59\% with GPT-3.5-turbo-0125.
HDLAgent offers an LLM-neutral framework to accelerate the adoption and growth
of HDL user bases in the era of agentic LLMs.


\end{abstract}

\section{Introduction}
\label{sec:Intro}

Recent advancements in Large Language Models (LLMs), such as OpenAI's GPT,
Google's Gemini, and Mistral AI's Mixtral are transforming the programming landscape. 
These models assist newcomers by providing intelligent
assistance, generating code snippets, and offering context-aware suggestions,
thereby significantly lowering the barriers to entry into programming.

However, LLMs currently offer limited support for niche Hardware Description 
Languages (HDLs), which often form specialized communities. Despite the 
ubiquity of Verilog\footnote{The original Verilog was designed in 1983, and modern versions like System-Verilog
are semantically compatible with it.}, emerging languages like Chisel3~\cite{chisel}, 
PyRTL~\cite{clow2017pyrtl}, and DSLX~\cite{dslx} illustrate the need 
for LLMs to adapt to new HDLs. The lack of training data for these 
languages means that existing LLMs underperform, creating a disincentive for developing new HDLs.

Additionally, many high-performance LLMs are closed-source, limiting 
their adaptability. Enhancing LLM capabilities without waiting for 
lengthy training cycles is crucial. Effective integration of LLMs 
with emerging HDLs would not only facilitate their adoption but 
also prevent LLMs from becoming barriers to innovation in hardware design.


To address these challenges, we propose a new LLM-neutral AI agent, {\bf HDLAgent}, that incorporates
state-of-the-art AI coding agent techniques, specifically adapted to support
multiple HDLs with limited LLM support. This addresses the challenge of
generating accurate and functional code in HDLs that have proven difficult for
existing LLMs.

AI agents~\cite{zhou2023agents} typically involve multiple workflow steps, 
including LLM prompts and interactions with external tools. 
Techniques such as self-reflection like Chain-of-Thought (CoT)~\cite{wei2023chainofthought}, 
memory enhancement through Retrieval Augmented Generation (RAG), 
and error minimization through grounding are commonly utilized. 
These techniques help improve the responses of LLMs by refining their input data and correcting syntax/semantics errors.


Although AI agents share these common workflow steps, they need to be adapted
to the specific problem. HDLAgent incorporates state-of-the-art AI coding agent
techniques in self-reflection and grounding using compiler errors, though these
concepts are similar to existing coding LLM works. The novelty in HDLAgent lies
in its approach to memory steps. We propose:
\begin{itemize}
\item Creating an HDL description summary to enable knowledge transfer between Verilog and the new HDL.
\item Generating few-shot learning examples that enrich the HDL description summary.
\item Enhancing grounding messages from compile errors to rectify them.
\end{itemize}

Additional contributions are the LLM performance evaluation across multiple
HDLs, and propose changes in the HDLs to better support LLMs.


HDLAgent succeeds where plain LLMs consistently fail. For instance, using
HDLAgent with \mixmoe yields a 44\% success rate when writing Chisel, compared
to just 3\% of tests passing without HDLAgent. Other LLMs, like \gpto, improve
from a 3\% success rate for DSLX to 48\% with HDLAgent. HDLAgent also benefits
LLMs with Verilog; for \mixlarge, the success rate increases from 13\% to 53\%.

In summary, the key contributions of this paper are:

\begin{itemize}
    \item \textbf{HDLAgent Development:} Introduction of HDLAgent, enhancing LLM performance in code generation for underrepresented HDLs.

    \item \textbf{Comprehensive Evaluation:} Detailed evaluations show HDLAgent boosts Chisel code success from 3\% to 44\%, and over 90\% for 
concise code samples across all HDLs.

    \item \textbf{HDL Enhancement Proposals:} Strategic modifications to HDL designs are suggested to guide future developments in HDL and compiler technologies.

    \item \textbf{Practical Impact and Adoption:} Our approach bridges significant gaps in applying LLMs to hardware design, 
simplifying the adoption of new HDLs and boosting developer productivity.
\end{itemize}


\section{Related Work}
\label{sec:related}

To adapt to a new language, there are two potentially complementary approaches
to improve LLM output: fine-tuning and Agents. These techniques can be
iteratively combined to develop Agents that produce even better results.

Fine-tuning is the process of adjusting the parameters of an LLM on a specific
dataset or task to improve its performance. Thus, fine-tuning can be applied to
optimize an LLM for a new language. RTLCoder~\cite{liu2024rtlcoder} fine-tunes
a 7B Mistral model with GPT-generated synthetic Verilog data. In contrast,
HDLAgent uses off-the-shelf LLMs without fine-tuning. The advantage of avoiding
fine-tuning is that many commercial flows do not allow it, and it is not a
trivial problem for languages with a very small set of examples.

URIAL~\cite{lin2023unlocking} bypasses the need for fine-tuning by enriching
prompts with illustrative examples. These prompts resemble the few-shot format
used by HDLAgent, incorporating both format and examples. While URIAL has shown
effectiveness in circumventing the need for instruction alignment, HDLAgent
further illustrates the possibility of learning previously unknown languages.


Agents~\cite{zhou2023agents} iterate through LLMs using three main techniques
to improve performance: self-reflection, memory, and grounding.


Self-reflection techniques use a sequence of interactions with the LLM instead
of a simple question/answer format. In this work, we call self-reflection to
chaining LLMs prompts to other LLMs. CoT~\cite{wei2023chainofthought} is an
example of self-reflection. Lumos~\cite{yin2023lumos} uses CoT to enable
simpler LLMs to outperform more advanced ones. These studies highlight
significant progress in this rapidly evolving field. Recent
works~\cite{yang2023large} propose an optimization method to find the best
prompt.


Memory techniques such as few-shot in-context learning~\cite{liu2022few,ahmed2022few}
and RAG~\cite{rag} use instructions, supplemental information, and
relevant examples to enhance efficiency. Various
methods exist for constructing prompts with extended context.
One such technique, querying an embedding database to augment the
context, is known as Retrieval Augmented Generation (RAG).


Grounding involves verifying or checking the LLM's response using an external
tool. While this is not always feasible, in code generation, a compiler or
testbench can validate and identify issues with the LLM-generated response.
This may trigger further iteration with the LLM.


Agents with self-reflection, memory, and grounding have been applied to improve
code generation. If we ignore the HDL target, and focus on generic programming
languages like Python or C++, several
works~\cite{tambon2024bugs,zhong2024chatgpt} show that errors can be fixed
by grounding the generated code against compiler or testbench feedback.
Intervenor~\cite{wang2023intervenor} proposes an Agent that successfully leverages compiler feedback. Other recent
works~\cite{moon2023coffee,dong2023selfcollaboration,wang2023intervenor,madaan2023self,ni2023lever,xia2023conversational,zhang-etal-2023-self,fan2022automated,olausson2023selfrepair}
propose Agents to iterate over testbench results to fix sematic errors in code.


VerilogCoder~\cite{zhao2024verilogcoder} introduces an autonomous coding approach using graph-based planning for synthesizing Verilog code from specifications. It combines a traditional LLM with a novel AST-based waveform tracing tool to refine the generated code. This tool traces the expected signal flow within the AST representation of Verilog, improving both the accuracy and reliability of the generated code. VerilogCoder has been shown to significantly reduce errors in synthesizable code by anticipating and correcting logical flaws before compilation.


The paper "Towards LLM-Powered Verilog RTL Assistant: Self-Verification and Self-Correction"~\cite{kumar2024llmpowered} discusses a framework that employs LLMs to not only generate but also verify and correct RTL designs in Verilog. This system uses a self-verification method that incorporates runtime feedback to iteratively refine the Verilog code, effectively decreasing the cycle time between code generation and testing. It represents a shift towards fully autonomous RTL design, promising to streamline the Verilog development process significantly.

Besides CoT, some notable self-reflection techniques for code generation include:
Self-planning~\cite{jiang2023selfplanning} proposes a planning stage or
self-reflection before code generation; Self-Debug~\cite{chen2023teaching}
proposes how to improve code generation by generating explanations in the
intermediate steps; ChatCoder~\cite{wang2023chatcoder} uses self-reflection to
paraphrase and elaborate on the initial question.


Early work~\cite{thakur2023verigen,10137086,yang2023new} with LLMs and Verilog
avoids using Agents because LLMs like \gptt are already reasonably trained for
Verilog. Several AI-based chip design competitions~\cite{efabless1,efabless2}
required designs implemented in Verilog. Looking at the top performers, they
tend to use \gptt and focus on combinational modules where the top level module
IO is fully specified. In all the cases, the human-in-the-loop guides the LLM
to fix problems with the generated code and iterate over the testbench results.


The same AI coding Agent concepts of self-reflection, memory, and grounding can be
applied to Verilog. Concurrent works include
AutoChip~\cite{thakur2023autochip}, RTLFixer~\cite{tsai2024rtlfixer}, and
HDLDebugger~\cite{yao2024hdldebugger}.

AutoChip~\cite{thakur2023autochip} uses testbench feedback to ground the
generated Verilog. It is similar to Self-Edit~\cite{zhang-etal-2023-self} and
Self-Repair~\cite{olausson2023selfrepair}, but with a focus on Verilog.
AutoChip focuses on simulation errors, which is not included in HDLAgent but is
a potential extension.

RTLFixer~\cite{tsai2024rtlfixer} uses ReAct~\cite{yao2023react} for
self-reflection and compiler errors for grounding. RTLFixer utilizes
human-generated explanations for various error messages, whereas HDLAgent uses
previous examples of errors and their respective fixes. Unlike RTLFixer, which targets only
Verilog, HDLAgent provides different error/fix strategies for each HDL.

HDLDebugger~\cite{yao2024hdldebugger} fine-tunes CodeLlama to fix code
generation, rather than to generate better Verilog, as RTLCoder does.
HDLDebugger uses compiler error messages to ground the generation and applies
this to the fine-tuned CodeLlama to fix the code. HDLDebugger represents a
different approach that, when available (publication expected in August 2024),
could be applied to HDLAgent for fixing compiler errors. However, it will
require fine-tuning for each HDL. From the provisional paper, HDLDebugger does
not seem to apply self-reflection.


For benchmarking, we use HDLEval~\cite{hdleval} and
VerilogEval~\cite{verilogeval} when possible. Both HDLEval and VerilogEval
include several tests derived from HDLBits~\cite{hdlbits}. HDLBits is a website
with problems and tests designed to teach students the basics of Verilog. These
tests are simple and exemplify the types of questions a person learning a new
HDL might have. Examples include outputs with a few lines of code, such as how
to rotate an input value.

HDLEval~\cite{hdleval} includes HDL-neutral tests, incorporating simple tests
from HDLBits as well as tests from the Efabless LLM competition. Importantly,
it categorizes tests into combinational and pipelining. This distinction is
crucial because some languages, like DSLX, do not support unrestricted
pipelining, allowing only combinational tests to be used.

VerilogEval~\cite{verilogeval} and RTLLM~\cite{lu2023rtllm} propose test sets
to evaluate only Verilog designs. VerilogEval consists of two sets of problems:
Human and Machine. The Human category includes tests generated by humans, while
the Machine category comprises tests translated into English from existing
Verilog code using GPT-3.5. RTLLM features a different set of problems divided
into arithmetic and logic. Both works use simulation for testing correctness
and evaluate only Verilog code. RTLLM claims that the tests could be used for
languages like Chisel, but the paper lacks explanations as how to address
issues with matching Chisel-generated IOs.

\section{HDLAgent}
\label{sec:hdlagent}

HDLAgent is an AI Agent (refer to Section~\ref{sec:related}) specifically tailored for adapting cutting-edge AI coding techniques to Hardware Description Languages (HDLs). This adaptation is crucial for HDLs that are not typically included in the training data of Large Language Models (LLMs).

LLMs demonstrate proficiency in transfer learning~\cite{pan2010survey,zhao2023expel}.
HDLAgent exploits this capability, enabling LLMs to handle HDLs with limited
training data. By facilitating the transfer of knowledge from well-known HDLs,
such as Verilog, to new HDLs such as PyRTL, HDLAgent empowers the LLM to adapt its understanding
of familiar programming languages to target languages. This process mirrors human
learning mechanisms~\cite{scholtz1989study}.

\begin{figure}[htbp!]
\begin{center}
\includegraphics[width=0.49\textwidth]{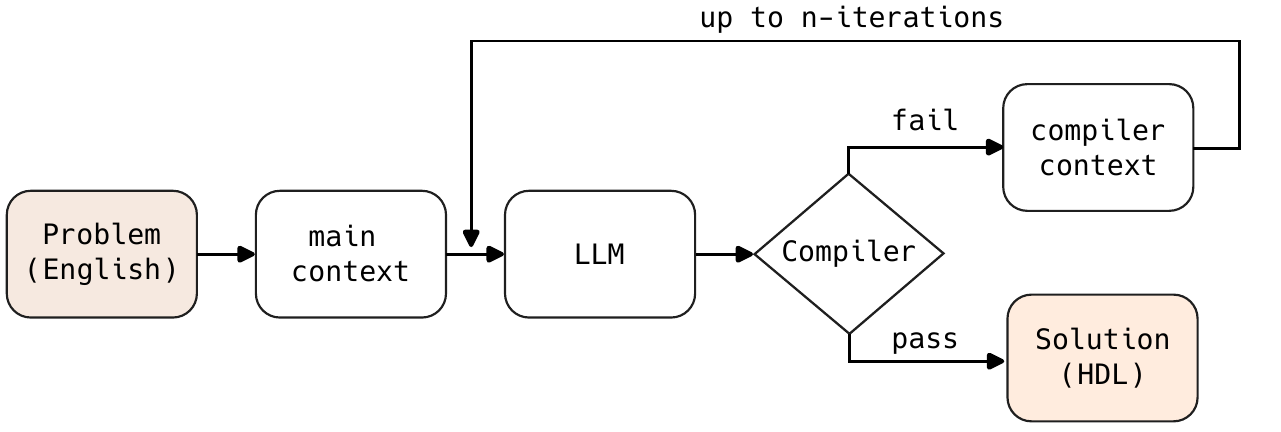}
\end{center}
\caption{HDLAgent flow leveraging compiler feedback.\label{fig:hdlagent}}
\end{figure}

As illustrated in Figure~\ref{fig:hdlagent}, HDLAgent addresses the critical 
challenge of limited HDL-specific knowledge. It employs two primary memory 
components to bridge this gap: the "main context" and the "compiler context". 
The "main context" (described in Section~\ref{sec:maincontext}) offers 
a succinct summary of HDLs along with targeted examples. The "compiler 
context" (Section~\ref{sec:compilercontext}) enhances code generation by 
integrating compiler feedback, grounding the output within practical and 
executable constraints.

\begin{figure}[htbp]
\begin{center}
\includegraphics[width=0.2\textwidth]{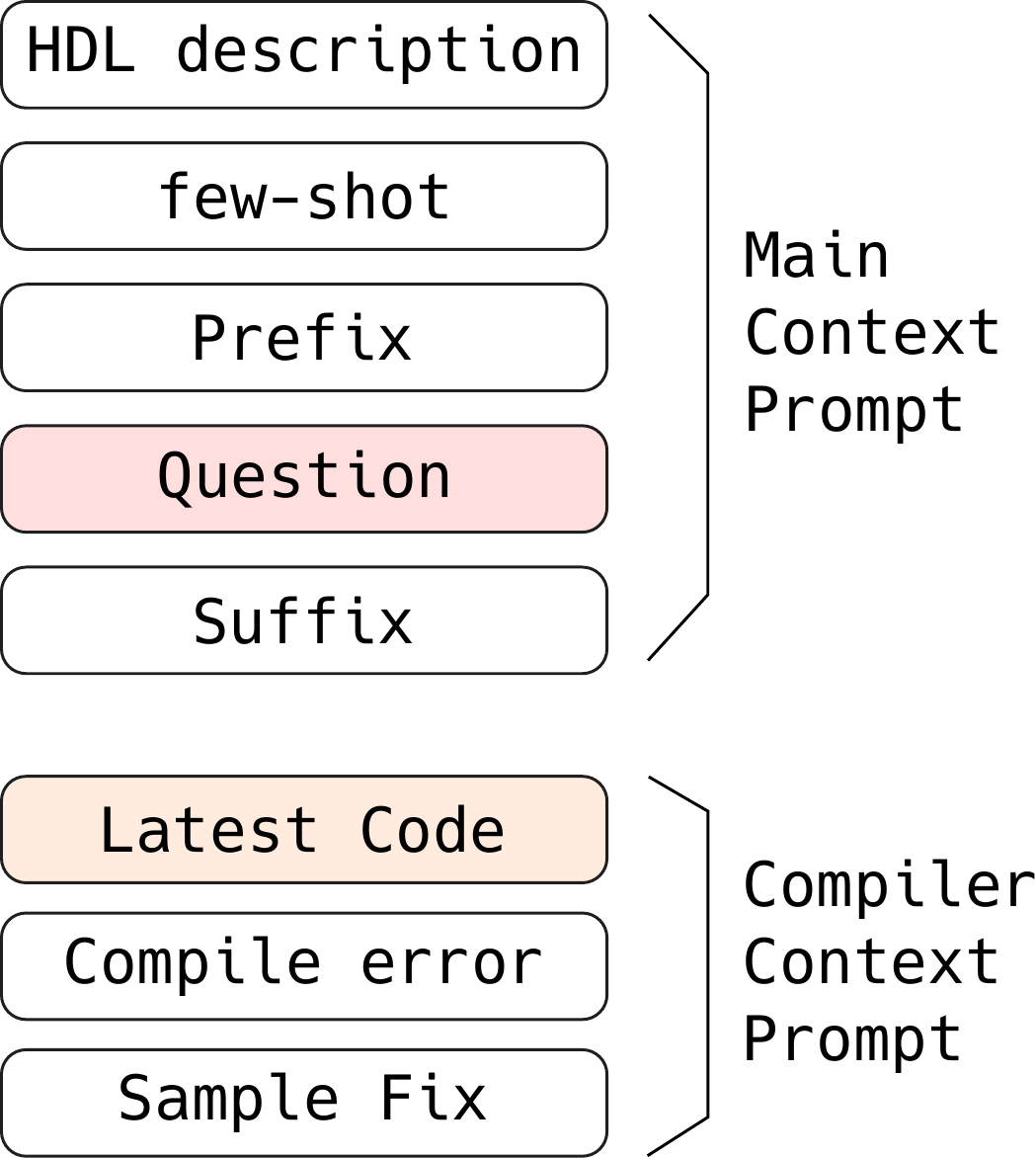}
\end{center}
  \caption{HDLAgent Main and Compiler context prompt components.}
\label{fig:context}
\end{figure}

\subsection{Main Context}
\label{sec:maincontext}

The "main context" in HDLAgent serves to inform the LLM about the specific HDL in use.
Figure~\ref{fig:context} illustrates this main context, which comprises four
key elements: HDL description, few-shot examples, Prefix, and Suffix.

The HDL description provides a concise summary of the HDL, tailored to the LLM's
familiarity with the language. While our evaluation demonstrates that the HDL
description can be optimized for each LLM, we opt for simplicity by selecting the
description that performs best with \mixmoe and \gptn. This choice is motivated by
these LLMs' lower proficiency in HDLs such as Chisel, DSLX, and PyRTL. It is worth
noting that the HDL description proves less beneficial only when the LLM already
excels in a given HDL, such as Verilog. Such few-shot are crucial, especially
in HDLs with unique syntaxes, helping LLMs avoid common pitfalls.

The prefix and suffix in the main context serve as navigational aids for the
LLM, directing the model's attention to the task at hand and setting boundaries
for its output. The prefix introduces the problem in the HDL’s language, while
the suffix provides specific instructions to ensure the output adheres strictly
to HDL syntax, avoiding unnecessary English explanations and maintaining
consistency in output formats.

Interestingly, even for LLMs capable of processing entire HDL reference manuals
within their context window, utilizing a summary enhances both success rate and
token efficiency. Our evaluation clearly demonstrates a significant improvement
in success rates when employing an HDL description. The intuition is that focusing
on the Verilog differences is more important than providing a lengthy description
of the language.

Since including a complete tutorial is neither practical nor advantageous
for the evaluated LLMs, we use an HDL description summary instead. To generate
these summaries, we leverage LLMs with large context windows, specifically \gptt
and \gpro, to condense the HDL reference manuals.

For PyRTL and Chisel, our evaluation revealed that the most effective prompt was
generated by \gptt using the following instruction: "PyRTL is a Hardware
Description Language with the following reference documentation and tutorial.
Create documentation useful for LLMs trying to generate PyRTL code. The generated
documentation should include code snippets and highlight any language syntax that
is atypical for HDLs."

For DSLX, the optimal summary was produced by \gpro using a similar prompt, with
the addition of "Be concise and avoid examples with similar syntax." at the end.
This minor variation in the prompt yielded the best results.

Complementing the HDL description, the "main context" provides few-shot
examples to illustrate common HDL operations and potential areas of confusion.
These examples cover bit operations, reductions, loops, multiplexing, and a
multiply-add block. While the HDL Description can include some examples, it
is important to cover these basic operations with simple examples, as LLMs
tend to revert to incorrect syntax.

The bit operations example demonstrates simple bit manipulation and concatenation,
while the reduction example showcases a basic NOR reduction over a given input.

Loops can be particularly confusing in some HDLs. For instance, in DSLX, all
variables are immutable, but loops have a special syntax for accumulator variables.
Including an example like the one in Listing~\ref{lst:dslx} in the HDL context
helps address cases where the LLM needs to create a loop.

The Prefix follows the few-shot examples, briefly directing the original
Question with a statement such as the following for DSLX: "The following
statements describe the problem to be addressed in DSLX."

The Suffix, appended after the question, serves to limit the scope of the problem
and provide specific instructions. It includes directives like "respond with valid
program syntax only, without additional English explanations" and tailors HDL input
and output formats. Handling I/Os is crucial, especially for HDLs with multiple
output options, necessitating instructions to maintain output integrity and naming
consistency. For DSLX, the Suffix includes directives like "do not split the outputs
into individual bits" and "variables assigned to the output struct should have the
same name as the struct fields." This Suffix concept is essential even when using
Verilog, as it can employ interfaces, structs, or plain Verilog-2001 syntax.

The HDLAgent Suffix facilitates interfacing between different HDLs and Verilog.
However, maintaining consistent naming conventions and common syntax remains a
critical issue that the Suffix must address, even in single-HDL use cases.

\subsection{Compiler Context}
\label{sec:compilercontext}

HDLAgent's compiler context employs an iterative approach to rectify
inaccurately generated HDL code. This process grounds the LLM-generated code by
providing feedback on potential errors or hallucinations.
Before submitting the LLM output to the compiler,
HDLAgent identifies the code section. This step is crucial, as LLMs may
generate English explanations despite explicit instructions to avoid them.

When the generated program fails compilation, producing a compiler error, HDLAgent
constructs a query (illustrated in Figure~\ref{fig:context}). This query begins
with the "main context," disregarding any non-code responses. It then incorporates
the latest code snippet, followed by a statement indicating "the previous code has
the following compile error," succeeded by the specific compiler error message. If
HDLAgent possesses an example fix for addressing the compiler error, it appends
this "sample fix" to the context.

The sample fix methodology is analogous to RTLFixer~\cite{tsai2024rtlfixer},
which provides explanations for resolving Verilog error messages. HDLAgent extends
this concept to cover multiple HDLs, elucidating the special syntax requirements
of a given HDL when necessary.

HDLAgent presents the entire latest code snippet in its query. We experimented
with a method inspired by CWhy~\cite{cwhy}, which focuses on a few lines of code
surrounding the compiler error message. While this approach worked for some
LLMs like \gptt, it proved less effective with others. Although this delta approach
reduces token usage, it led to increased error rates, prompting us to exclude it
from our evaluation. As LLMs continue to evolve, this approach may warrant
reconsideration in future iterations.

\subsection{Prompt Optimizations}

Besides the previous main and compiler context there are also several subtle but important optimizations:

\begin{itemize}
  \item Placing the prompt after the context achieves better results~\cite{islam2023financebench}.

  \item HDLAgent approach avoids the chat-like history with all the previous
    code generations and fixes. Keeping the original question iteration but not
    the compiler error fixes achieves better results~\cite{thakur2023autochip}.
    We did a quick test with HDLAgent and DSLX. Avoiding a history with all the
    error fixes had a 5\% improvement in \gptt and a 27\% in \mixmoe.

  \item Most LLMs generate code snippets in quoted sections, but not always.
    Even worse, it is common to write English explanations even thought the
    prompt explicitly asks to just write code. To address this, for each
    language we have a filter/detector that removes English and finds code
    boundaries. For example in Verilog it allows preprocessor directives and
    code between module and endmodule. Without this, some smaller LLMs fail
    very frequently.

\end{itemize}

\subsection{LLM Cost}
\label{sec:llmcost}

Our approach approximates LLM cost by the number of tokens utilized,
serving as a practical proxy for monetary cost and compute resources
required. As context length increases, so do the costs and computational demands.

While token 
usage offers a simple metric for gauging efficiency, our primary focus remains 
on balancing accuracy with cost-efficiency. This approach necessitates a 
judicious use of context and iteration, ensuring that each interaction 
with the LLM is as productive as possible. Future studies
could investigate efficiency metrics like ${\text{error rate}} \times
{\text{tokens}}$.

A crucial aspect to consider is the "stateless" nature of LLMs, where each call
requires a complete context. APIs like OpenAI and Mixtral lack a "chat-like"
interface that accumulates context across queries, unlike \gpro, which can retain context.
Depending on the cost model for LLMs like \gpro, retaining history may be more efficient.
However, the context length of \gpro is insufficient for handling
multiple iterations. Therefore, in this work, we flush the history and disregard
cost models, considering only total token usage after all HDLAgent
iterations.

\section{Setup}
\label{sec:setup}

Table~\ref{tab:tools} lists all the languages used in the evaluation and the
compiler versions used by this paper. When a date is provided it corresponds to
the top-of-tree version at that given month. For Quality of Results (QoR), we
use Yosys synthesis results.

\begin{table}[ht]
\centering
\caption{Language Tools and Versions}
\label{tab:tools}
\begin{tabular}{l|l|l}
\toprule
\textbf{Language}         & \textbf{Tool}           & \textbf{Version} \\
\midrule
Verilog                   & Yosys                   & 0.35               \\
Chisel                    & FIRRTL                  & 3.5.0-RC2           \\
PyRTL                     & PyRTL compiler          & 0.10.2               \\
DSLX                      & XLS                     & 3/2024               \\ 
\bottomrule
\end{tabular}
\end{table}

\begin{table}[ht]
\centering
\caption{LLMs used in the evaluation}
\label{tab:llm}
\begin{tabular}{l|l|l|l}
\toprule
  \textbf{LLM}  & \textbf{Version}    & \textbf{Date}    & \textbf{Context}  \\ 
\midrule
  \gptt                 & gpt-4-1106-preview  & 4/23              & 128000 \\
  \gptn                 & gpt-3.5-turbo-0125  & 9/21              & 16385  \\
  \gpto                 & gpt-3.5-turbo-1106  & 9/21              & 16385  \\
  \gpro                 & gemini-1.0-pro-001  & 2/24              & 32720  \\
  \mixmoe               & Mixtral-8x7B-instruct   & 12/23             & 32768  \\
  \mixlarge             & Mixtral-8x22B-v0.1  & 3/24              & 32768  \\
\bottomrule
\end{tabular}
\end{table}

Table~\ref{tab:llm} shows the LLMs used. Many LLMs, including~\gpto , are not
deterministic. They have produced differing outcomes for the same example under
identical prompt conditions. OpenAI recently proposed a new API to address this
issue, providing a seed, but this solution still needs to be fully implemented
across all LLMs. For fair evaluation, we avoid the deterministic settings and
perform 1, 5, or 10 runs depending on the top@k parameter.



\section{Evaluation}
\label{sec:evaluation}

\subsection{Overall Results}
\label{sec:hdlagent_eval}

To comprehensively assess HDLAgent's performance across various LLMs, we evaluate
each HDL (Chisel, PyRTL, DSLX, and Verilog) against four benchmark tests: VH
(VerilogEval-Human), VM (VerilogEval-Machine), HC (HDLEval-Comb), and HP
(HDLEval-Pipe).

While VerilogEval tests comprise several Verilog-specific questions, they do not
fully demonstrate the potential of the LLM/HDLAgent combination as effectively as
HDLEval (HC, HP). This is primarily due to VerilogEval's inclusion of
Verilog-specific instructions in some tests, such as implementing a D latch using
an "always" block. Such tests are not suitable for evaluating languages other than
Verilog. Consequently, we utilize VH and VM primarily for Verilog or as a reference
point, while focusing our evaluation on HDLEval (HC and HP).

The results are broken down into five key components to delineate the incremental benefits provided by HDLAgent:

\begin{itemize}
  \item \textit{Base}: Represents the baseline performance of an LLM without 
      HDLAgent enhancements but includes basic I/O formatting and general code generation guidelines.

  \item \textit{Description}: Adds a concise HDL Description to the LLM context, 
      improving specificity (see Section~\ref{sec:maincontext} for details).

  \item \textit{Few-shot}: Adds language-specific few-shot examples.
    Section~\ref{sec:contextinsights} provides further insights on the HDL
    Description and few-shot context selection.

  \item \textit{Compile}: Incorporates compiler feedback with up to eight 
    iterations to refine the generated code, optimizing accuracy 
    (justification for the number of iterations is in Section~\ref{sec:passsensitivity}).

  \item \textit{Fixes}: Performs the same iterations as \textit{Compile}, but
    for each iteration, provides a suggestion alongside a generic example on
    how to address the specific compiler error.
\end{itemize}

\begin{figure}[htbp]
\begin{center}
\includegraphics[width=0.45\textwidth]{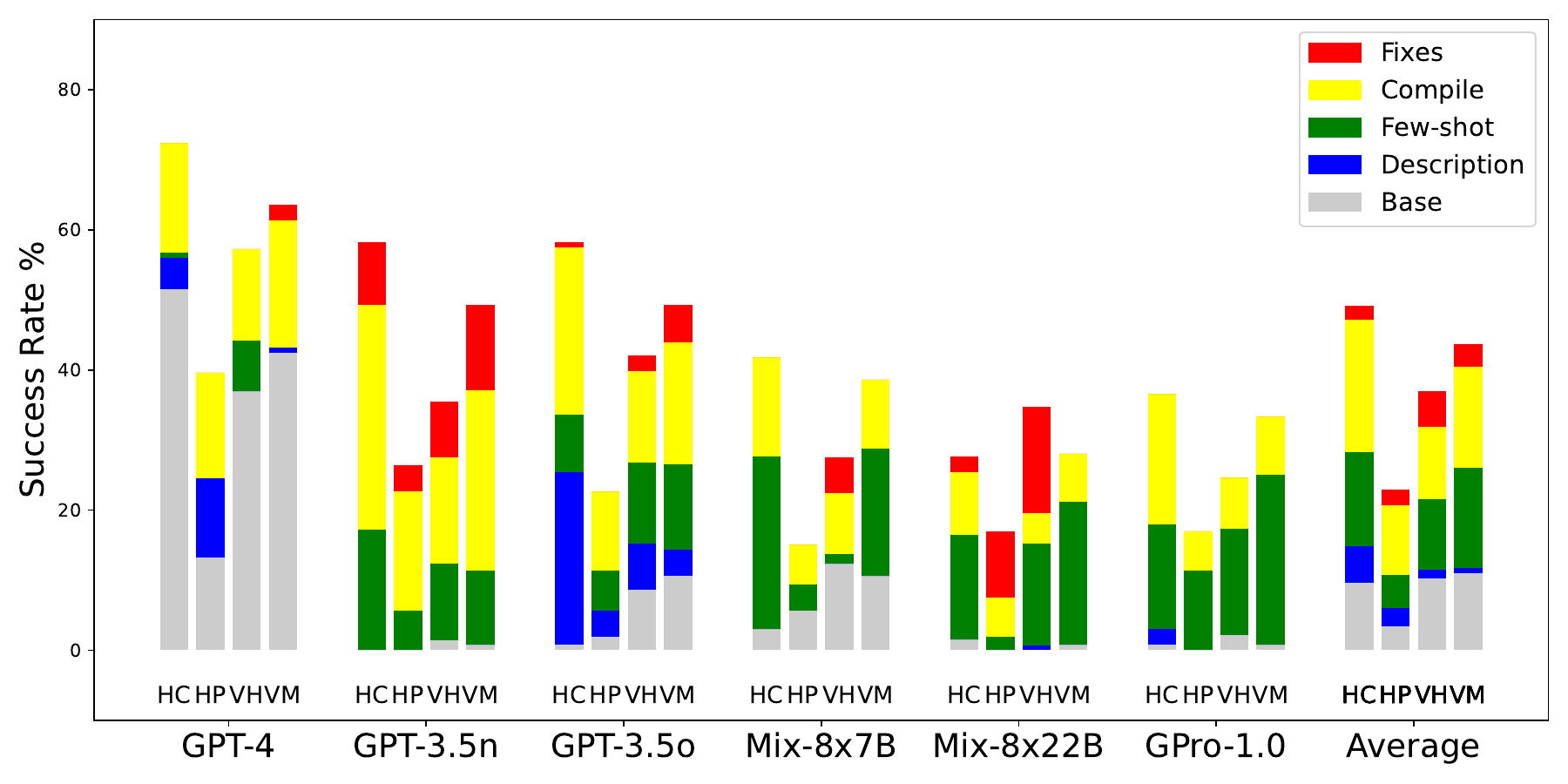}
\end{center}
\caption{HDLAgent improves Chisel across all LLMs.\label{fig:chisel}}
\end{figure}

{\noindent \bf Chisel} (Figure~\ref{fig:chisel}), a Scala-based HDL, presents a
unique challenge and opportunity. Most LLMs are familiar with Scala but have limited knowledge of Chisel.
While several LLMs demonstrate familiarity with its basic
syntax, only \gptt initially performs adequately with Chisel (52\% success
rate). All other LLMs exhibit a mere 3\% success rate or less. Both the "main
context" (comprising Description and Few-shot components) and the "compiler
context" (including Compile and Fixes elements) provide substantial benefits,
underscoring the necessity of all these components. Notably, with HDLAgent,
\gpto and \gptn outperform even high-performing LLMs like \gptt in its baseline
state. Furthermore, HDLAgent significantly enhances \gptt's performance,
elevating its success rate to 72\%.

Examining the average performance across all LLMs reveals that each component
of HDLAgent contributes significantly to the overall improvement. This
underscores the comprehensive and synergistic nature of HDLAgent's approach to
enhancing LLM performance in Chisel code generation.

\begin{figure}[htbp]
\begin{center}
\includegraphics[width=0.45\textwidth]{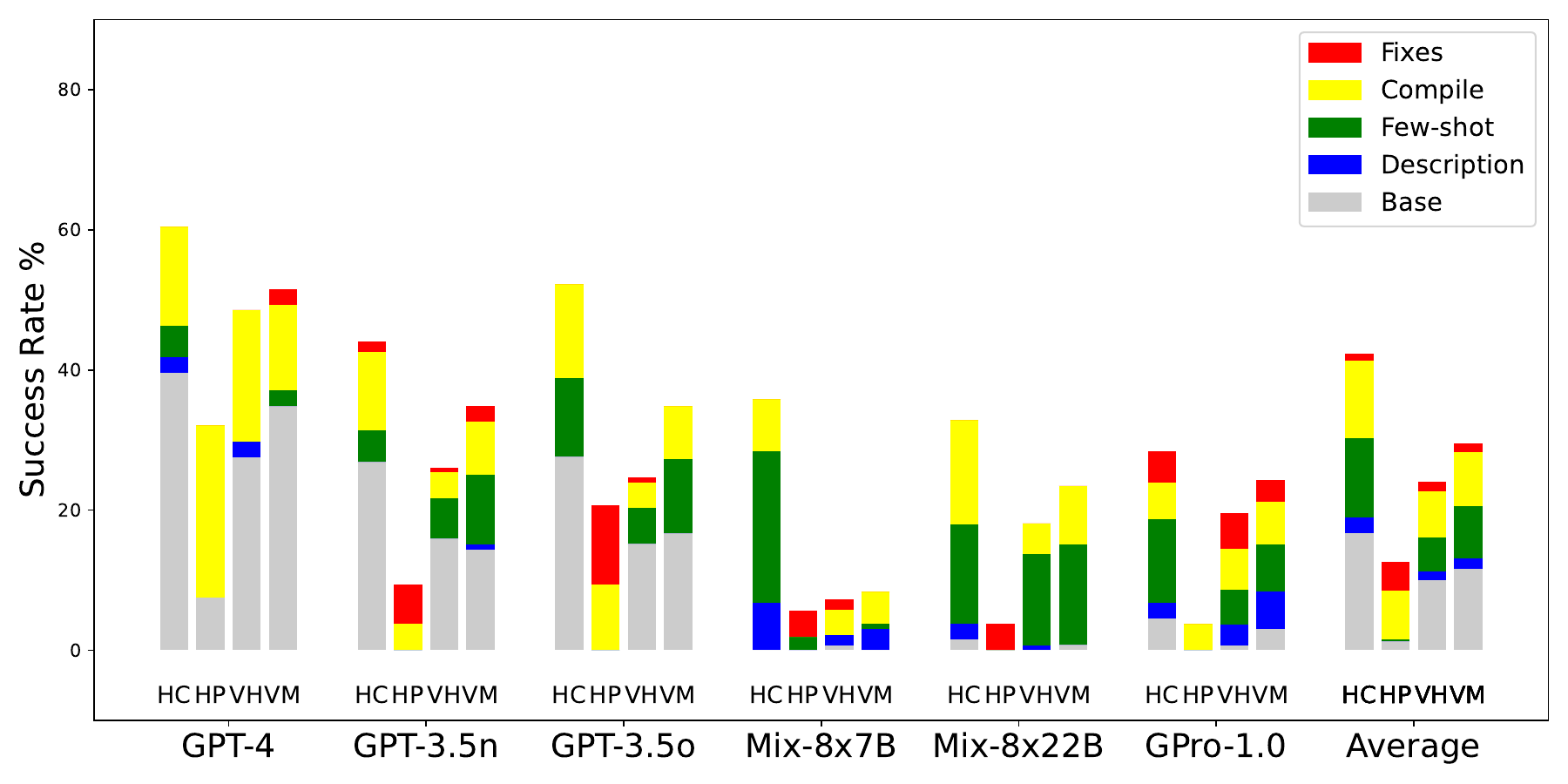}
\end{center}
\caption{HDLAgent improves PyRTL across all LLMs.\label{fig:pyrtl}}
\end{figure}

{\noindent \bf PyRTL} (Figure~\ref{fig:pyrtl}), a Python-based Domain Specific Language (DSL), presents
challenges similar to those of Chisel. OpenAI's LLMs (\gptt, \gptn, \gpto)
demonstrate some capability in passing several tests without HDLAgent (Base)
due to the strong baseline performance in Python; however, their success rates
for PyRTL remain low, ranging from 27\% to 40\%. When HDLAgent is implemented,
these success rates significantly improve, increasing to a range of 44\% to
60\%. As observed in the Chisel evaluation, HDLAgent's performance boost stems
from multiple factors.

Mirroring the Chisel results, all components of HDLAgent prove important for
PyRTL, with the "compiler context" (\textit{Compile} + \textit{Fixes}) playing
a particularly crucial role. This heightened importance of the compiler
iterations for both PyRTL and Chisel can be attributed to their nature as
DSLs and therefore their error message generation. Using the
PyRTL or Chisel compiler error messages, HDLAgent iterates to fix the code.
Since the baseline LLM knows the language, it can
interpret the compiler error messages and iterate to fix mistakes.

LLMs often confuse the syntax of DSL host languages (Python for PyRTL, Scala
for Chisel) with HDL-specific syntaxes. The HDL Description significantly aids
compiler iterations in rectifying these mistakes, demonstrating synergy between
the Description and Compile passes. Although not shown in the Figure, enabling
only the Compile pass without the HDL Description and Few-Shot examples yields
substantially lower overall improvement. This issue could be mitigated if PyRTL
or Chisel compilers generated errors that clarified the distinction between
base languages and DSLs, perhaps by providing a single few-shot example.

\begin{figure}[htbp]
\begin{center}
\includegraphics[width=0.45\textwidth]{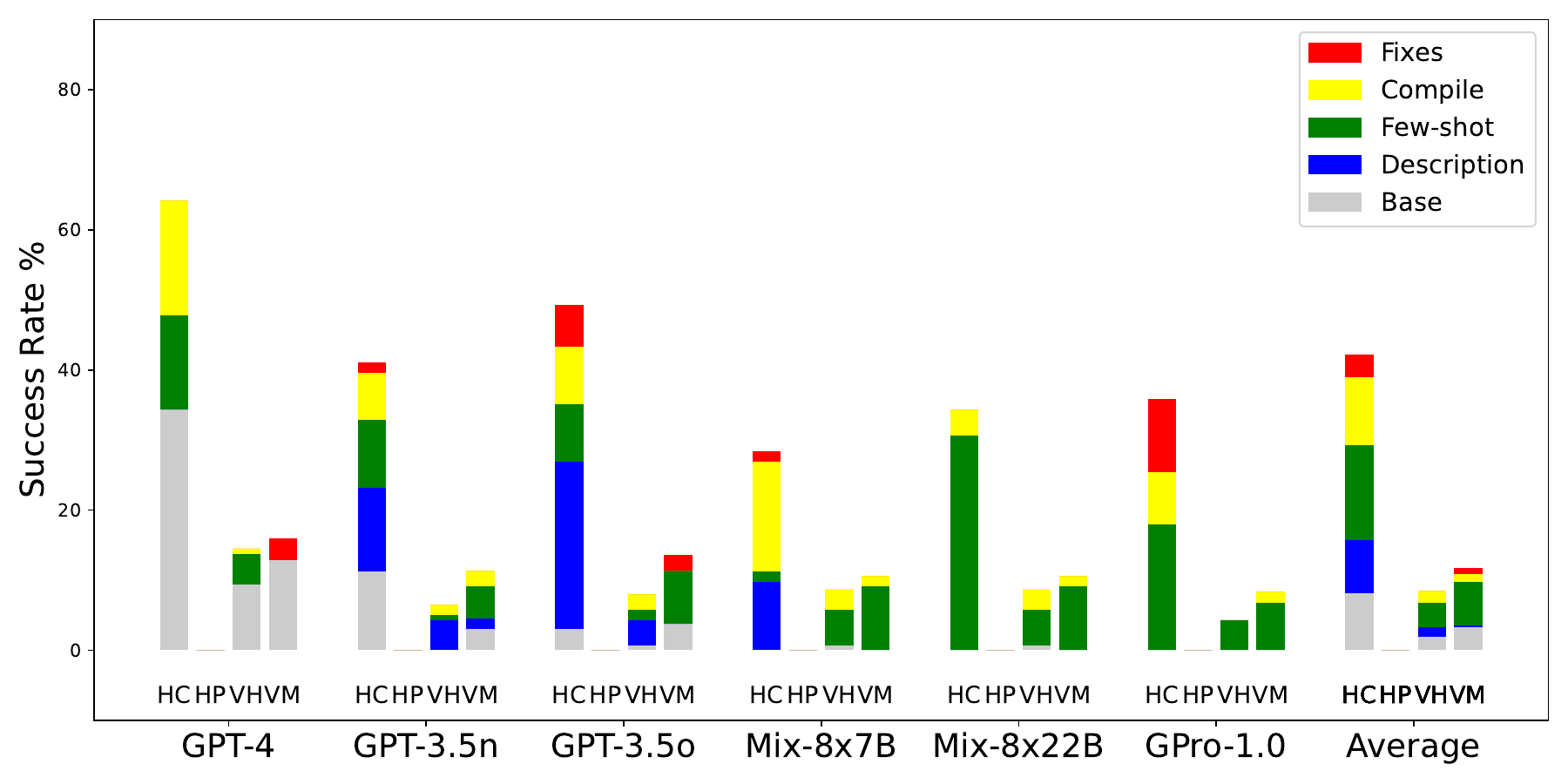}
\end{center}
\caption{HDLAgent improves DSLX HDLEval-Comb across all LLMs.\label{fig:dslx}}
\end{figure}

{\noindent \bf DSLX} (Figure~\ref{fig:dslx}), a Rust-like language, presents
unique challenges for implementing a Rust-like syntax that it is not fully
compatible with Rust. Due to its limitations on arbitrary pipelining, DSLX
cannot be evaluated against HDLEval-Pipe and performs poorly with VerilogEval.
While \gptt demonstrates some DSLX knowledge, HDLAgent significantly enhances
results across all LLMs.

Unlike Chisel and PyRTL, DSLX is not a DSL. Consequently, the "main context"
(\textit{HDL Description} + \textit{Few-Shot}) emerges as the primary factor in
HDLAgent's improvement. Explaining the Rust-like syntax and providing examples
proves more crucial than grounding results with compile errors.

This shift in importance from compiler feedback to Description stems from DSLX's
unique syntax. While it resembles Rust, not all Rust syntax is valid in DSLX. In
contrast, Chisel and PyRTL accept all Scala and Python syntax, respectively. Without
clear guidance on DSLX's specific syntax, LLMs struggle to generate correct code.

\begin{figure}[htbp]
\begin{center}
\includegraphics[width=0.45\textwidth]{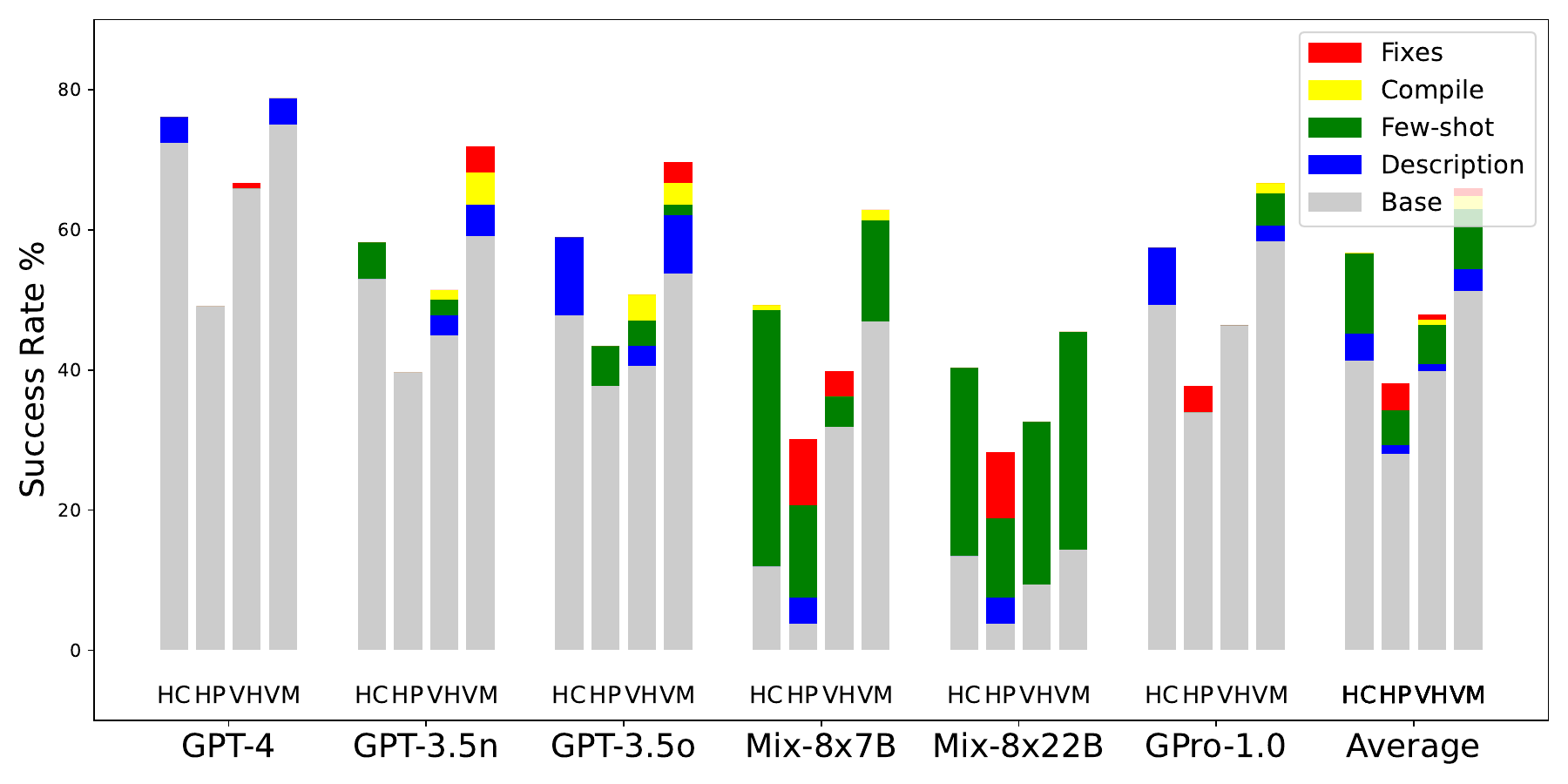}
\end{center}
\caption{Verilog succeeds across benchmarks and LLMs .\label{fig:verilog}}
\end{figure}

\begin{figure*}[htbp]
\begin{center}
\includegraphics[width=0.74\textwidth]{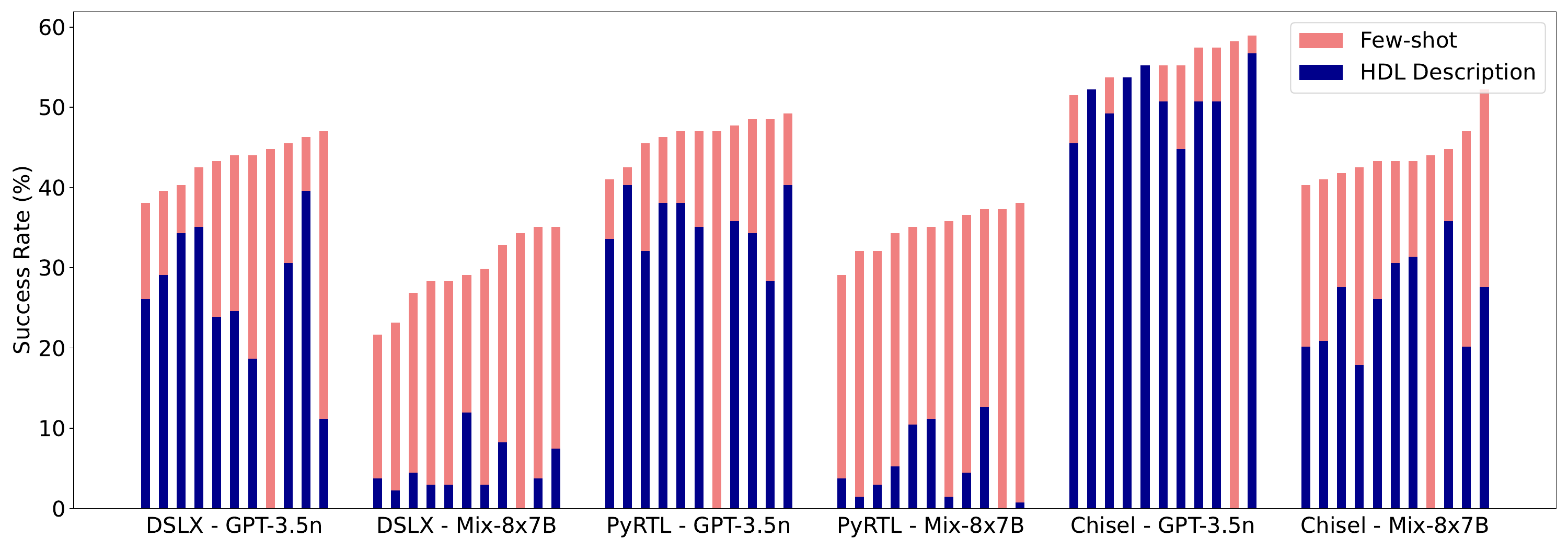}
\end{center}
\caption{HDL description and few-shot help LLMs to improve results.\label{fig:evalcontext}}
\end{figure*}

{\noindent \bf Verilog} (Figure~\ref{fig:verilog}) demonstrates the best overall
performance in the \textit{Base} condition (without HDLAgent), as expected due to
extensive training with Verilog syntax. It is also the only fair case for VerilogEval
use. HDLAgent minimally impacts models already proficient in Verilog but significantly
enhances \mixmoe and \mixlarge, which have some Verilog knowledge, illustrating
effective knowledge transfer even with limited Verilog familiarity.

Compiler iterations provide little benefit in error recovery for Verilog. This
is due to LLMs' higher proficiency in Verilog syntax. Manually analyzing the HC
results, only 5 out of 134 tests showed Verilog syntax errors, with just one
potentially benefiting from improved error messaging. As a result, using
Slang~\cite{slang} instead of Yosys~\cite{yosys} did not improve results for
\gpto despite producing more descriptive error messages.

Unexpectedly, \mixlarge underperforms \mixmoe, possibly due to difficulties in
following directions. An "instruct" model might yield better results, but we
retained the non-instruct model for its insights into HDLAgent's impact across
LLMs.

HDLAgent successfully enables LLMs to use new HDLs. Comparing \gpto and \gptn
across HDLs shows consistent relative performance regardless of the LLM used.
For example, with \gptt, Verilog achieves a 76\% success rate, while PyRTL, the
lowest, reaches 60\%. This pattern holds across all tested LLMs. Even the
worst-performing LLM (\mixlarge) achieves a 53\% success rate with Verilog and
28\% with PyRTL, a significant improvement from the zero success rates many
LLMs had without HDLAgent.


\subsection{HDLAgent Context Insights}

\label{sec:contextinsights}

%
%

This section offers insights into the selection of HDL Description and few-shot
context. One straightforward approach is to utilize the full reference manual
directly for the specific language. While this is feasible for models with
large context windows such as \gptt, \mixmoe, and \gpro, it generally proves
less effective than employing a summarized HDL description. For instance, using
a full reference instead of a summary yields no change in results for \gpro,
but reduces the success rate from 77\% to 66\% for \gptt, and from 59\% to 33\%
for \mixmoe. These findings indicate that future LLMs need to improve their
handling of lengthy contexts, as all evaluated models struggle with this
aspect. Nevertheless, even if the LLMs improve, it is still advantageous to use
smaller summary context because it reduces the LLM cost.

Figure~\ref{fig:evalcontext} shows the DSLX, PyRTL, and Chisel success rate as
different reference manuals are summarized for HDLAgent. Each bar shows a
different LLM reference summarization prompt (Section~\ref{sec:maincontext})
sorted by accuracy. The breakdown is the contribution of the few-shot examples
and the HDL description. Interestingly, adding Few-shot always improves
results, and removing HDL Description and just keeping few-shot examples is a
reasonable alternative. In some HDL/LLM combinations like Chisel/\gptn, using
either Few-shot or Description works. For other combinations like DSLX/\mixmoe,
HDL Description helps but Few-shot is necessary. Optimal results require both
Few-shot and HDL Description.

\subsection{Pass Sensitivity}
\label{sec:passsensitivity}

Top@k is a popular method that measures how results can be improved by generating
multiple attempts. A k=5 means that when 5 LLM tries are used, at least one has
the correct code generation. Table~\ref{tab:topk} shows tests passed for
HDLEval-Comb for multiple LLMs and multiple top@k values (1,5,10). Due to
space, only the HDLEval-Comb results are shown.

Less popular HDLs benefit more from higher top@k values. For example, DSLX
shows a 1.22 to 2.08 times improvement in test pass rates from top@1 to top@10.
Verilog has between 1.16 and 1.45 times. This discrepancy is likely because the
LLM, unfamiliar with the language, starts from an incorrect baseline and
struggles to correct errors through compiler feedback. Not being able to recover is
very rare in Verilog but over 10\% of the DSLX tests have this problem. The
higher the top@k, the easier it is to avoid. Once the code compiles correctly, the
failure rate for all the HDLs is comparable. This means that if a future
HDLAgent improved the iterations or selected better starting points, it could further
improve results.

\begin{table*}[htp!]
  \footnotesize
\centering
\begin{tabular}{ll|ccc|ccc|ccc|ccc|}
\toprule
  & & \multicolumn{3}{c}{Verilog} & \multicolumn{3}{c}{Chisel} & \multicolumn{3}{c}{PyRTL} & \multicolumn{3}{c}{DSLX} \\
\cmidrule(lr){3-5} \cmidrule(lr){6-8} \cmidrule(lr){9-11} \cmidrule(lr){12-14}
  & & k=1 & k=5 & k=10 & k=1 & k=5 & k=10 & k=1 & k=5 & k=10 & k=1 & k=5 & k=10 \\
\midrule
  \multirow{2}{*}{\gptt}
  & Base & 97 & 103& 111&69 &88 &92 &53 &79 &85 &46 &79 &85 \\
  & HDLAgent &102 & 109& 111& 97& 103& 107& 81&92 &98 &86 &100 &104 \\
  \midrule
  \multirow{2}{*}{\gptn}
  & Base &71 & 96& 100& 0&5 &9 &36 &63 &67 &15 &32 &41 \\
  & HDLAgent & 78& 93& 98&80 &97 &100 &59 &79 &88 &55 &80 &88 \\
  \midrule
  \multirow{2}{*}{\gpto}
  & Base &64 &93 &99 &1 &6 &14 &37 &60 &71 &4 &19 &25 \\
  & HDLAgent &79 &92 &100 &79 &91 &99 &70 &78 &89 &65 &86 &91 \\
  \midrule
  \multirow{2}{*}{\gpro}
  & Base &66 &97 &105 &1 &5 &12 &6 &17 &31 &0 &0 &0 \\
  & HDLAgent &77 &96 &99 &49 &84 &88 &38 &66 &77 &48 &74 &82 \\
  \midrule
  \multirow{2}{*}{\mixmoe}
  & Base &16 &39 &50 &4 &12 &17 &0 &1 &2 &0 &0 &0 \\
  & HDLAgent &66 &86 &95 &60 &80 &86 &48 &71 &82 &38 &72 &79 \\
  \midrule
  \multirow{2}{*}{\mixlarge}
  & Base &18 &65 &78 &2 &12 &18 &2 &8 &13 &0 &0 &6 \\
  & HDLAgent &72 &96 &101 &35 &79 &89 &39 &67 &72 &47 &75 &81 \\
\bottomrule
\end{tabular}
  \caption{\label{tab:topk} top@k results for HDLEval-Comb for different
LLMs with just a Base query or with HDLAgent.}
\end{table*}

\begin{figure}[htbp]
\begin{center}
\includegraphics[width=0.3\textwidth]{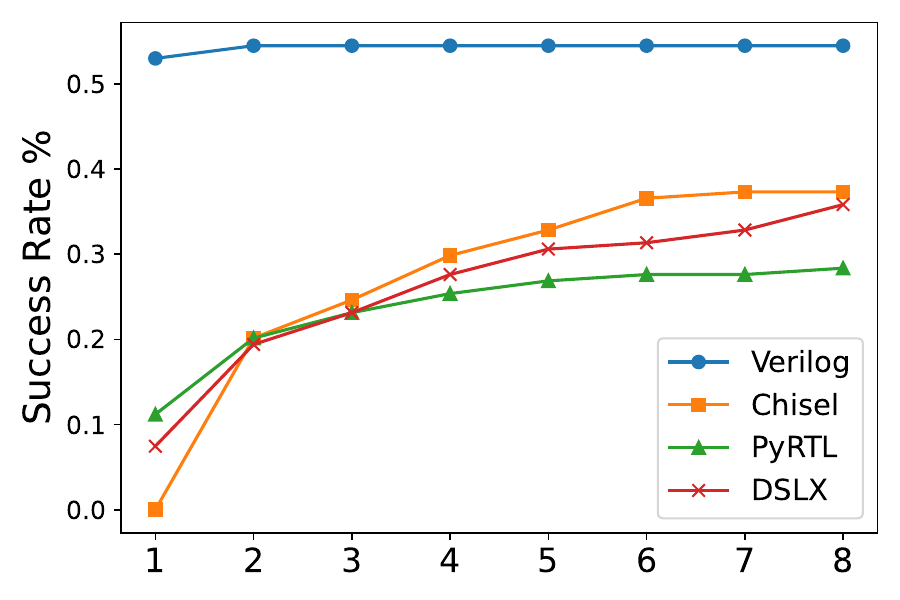}
\end{center}
\caption{\gpro converges in a few iterations.\label{fig:iterations}}
\end{figure}

Figure~\ref{fig:iterations} provides further insights into the top@k results,
illustrating the increase in accuracy as HDLAgent iterates with the compiler for
\gpro across various HDLs. We selected \gpro for this analysis as it requires more
iterations to converge compared to other LLMs. While Verilog converges rapidly,
other HDLs necessitate 6 to 8 iterations for convergence. Additional iterations
beyond this point do not improve the success rate, but altering the starting point,
such as using top@5, does enhance results. Overall, 8 iterations prove sufficient
across languages, as increasing iterations further fails to improve success rates
while incurring higher token usage.

When employing top@5 and 8 iterations (Table~\ref{tab:topk}), HDLAgent-supported
HDLs (Chisel, PyRTL, DSLX) perform equal to or better than the same LLM with Verilog
(Base). This finding represents a key contribution of the paper, demonstrating
that HDLAgent effectively enables the use of less popular, community-developed HDLs.

\subsection{Time and QoR}

\begin{figure}[htbp]
\begin{center}
\includegraphics[width=0.3\textwidth]{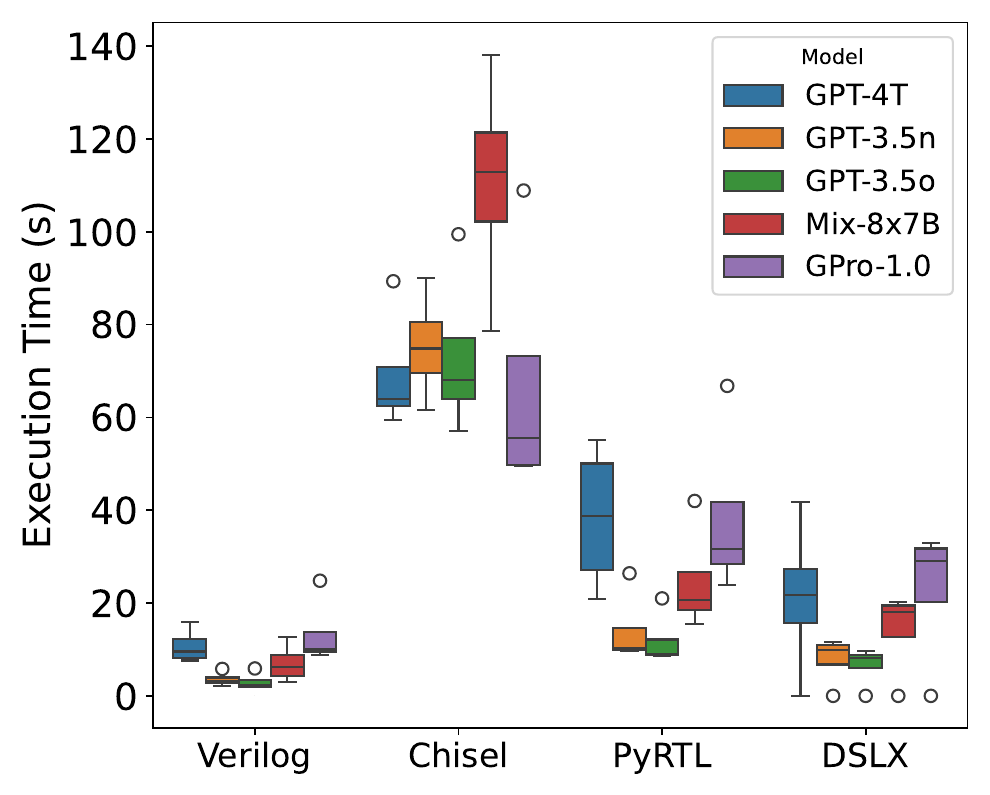}
\end{center}
\caption{LLM and HDL affect total HDLAgent execution time.\label{fig:time}}
\end{figure}

Execution time is a crucial metric for any AI Agent. It refers to the time HDLAgent
requires to generate a response, not the quality of results (QoR).
Figure~\ref{fig:time} presents a boxplot of execution times for HDLEval-Comb across
different LLMs, encompassing both successful and failed tests. All languages except
Verilog undergo a translation process to Verilog, adding overhead. In HDLAgent, the
execution time is a function of $\frac{\textit{tokens}}{\textit{second}}$, the number
of iterations, and external compiler speed.

Among the HDLs, Chisel stands out as the main outlier, with approximately 2/3 of
the execution time consumed by the FIRRTL compiler generating Verilog. \gptt exhibits
faster performance due to fewer errors and consequently fewer iterations. PyRTL and
DSLX also show slower performance than Verilog, partly due to additional
iterations.

Comparing LLMs, \gptn and \gpto generally demonstrate faster overall performance,
combining fewer error iterations with rapid result generation. External
$\frac{\textit{tokens}}{\textit{second}}$ benchmarking~\cite{aianalysis} indicates
that \gpro is approximately 30\% faster than \gptn and four times faster than \gptt.
However, HDLAgent results differ due to variations in iterations and speed.

\begin{figure}[htbp]
\begin{center}
\includegraphics[width=0.40\textwidth]{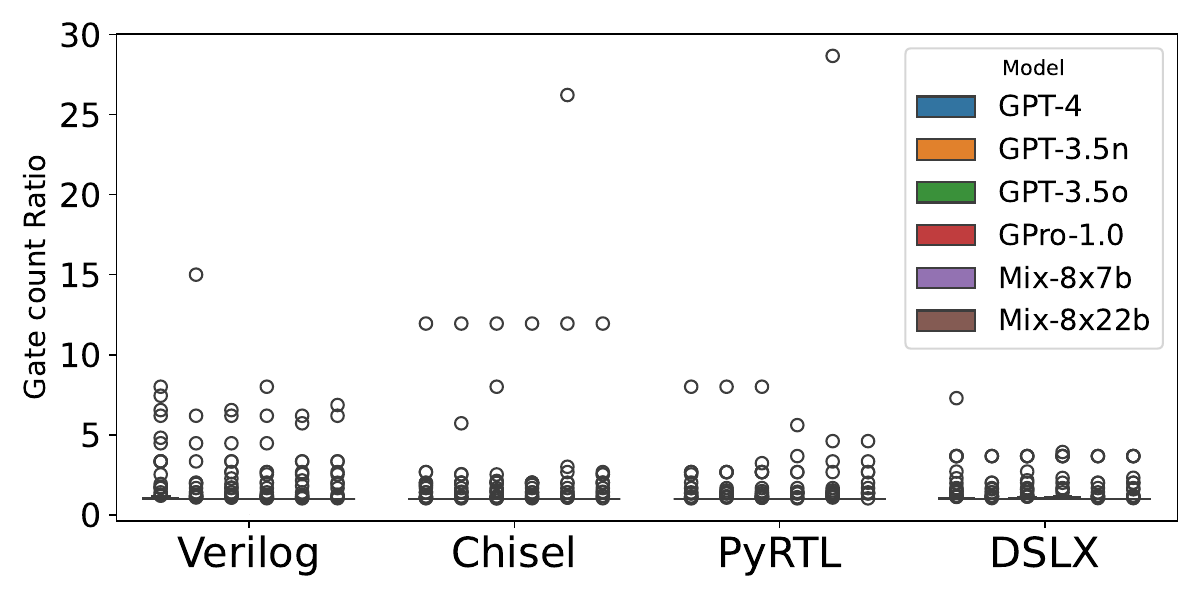}
\end{center}
\caption{QoR is consistent across LLMs but different across HDLs.\label{fig:qor}}
\end{figure}

Quality of Results (QoR) is paramount in hardware generation. The tests in HDLEval
are relatively small (under 500 LoC of equivalent Verilog), with many being purely
combinational. Consequently, frequency or power QoR metrics are not relevant. Instead,
we quantify QoR as the ratio of gates used compared to the best known implementation
for each module request.

Figure~\ref{fig:qor} illustrates the gate count ratio relative to the optimal
implementation. A ratio of 1 indicates optimal gate count, while 2 signifies
double the optimal count. An interesting observation is that for many designs,
we used the generated code as the optimal result.
The hand-generated reference Verilog for HDLEval often turned out to be less optimal.

Figure~\ref{fig:qor} only includes successful runs using HDLAgent with HDLEval-Comb. The
plot reveals significant QoR variation compared to the best implementations.
Typically, averages are skewed by one or two outliers. For example, in PyRTL
generated by \mixmoe, the average gate count ratio is 1.63, but drops to
1.12 when two outliers are removed. This suggests that LLMs occasionally
  generate highly inefficient code, but such instances are infrequent.

A second observation indicates that \gptt may appear to underperform; however, this
is partly due to its ability to successfully implement larger and more complex designs
that are difficult to optimize, which affects
the overall results. A third observation is that the efficiency of code generated
by various LLMs is generally comparable. Among these, DSLX appears to be the most
efficient, albeit by a slim margin. In DSLX generated by \gptt, 80\% of the produced code achieves the optimal
1:1 ratio. This suggests that an efficient compiler like XLS, combined with a popular
syntax, can yield superior results for generated HDL code.


\subsection{Usefulness Insights}
\label{sec:useful_eval}

HDLAgent significantly improves the LLM performance across all LLMs and HDLs, but
in some cases like DSLX, the average HDLEval-Comb performance is around 60\%. This
can be interpreted as not good enough because it fails many times.

This section provides more insights in which tests pass and fail. HDLEval-Comb comprises 134
tests, with some being relatively small, containing only a few lines of code, while others are
significantly more extensive. HDLEval is designed to encompass a range of tests, from straightforward
to complex.


\begin{figure}[htbp]
\begin{center}
\includegraphics[width=0.34\textwidth]{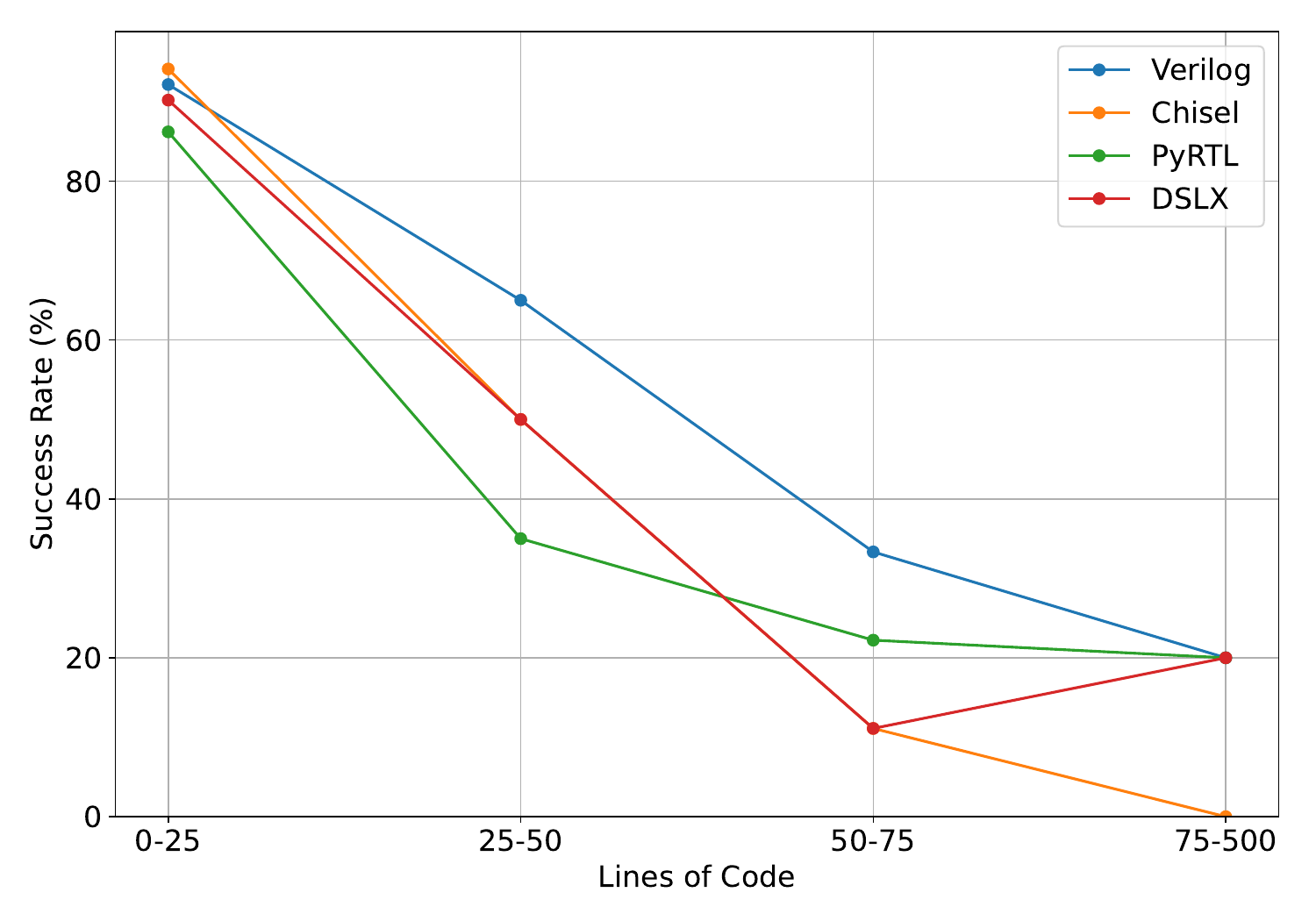}
\end{center}
  \caption{Even with best LLM (\gptt), performance degrades as Lines of Code for generated output increases.\label{fig:locoverall}}
\end{figure}

The output program complexity provides key insights in current LLM and HDLAgent
limitations. The best proxy for complexity is not the input problem itself, but
the lines of code (LoC) required to implement such a problem in a specific
language (Verilog in this case).

Figure~\ref{fig:locoverall} illustrates the success rate for HDLEval-Comb using
\gptt across four different problem sizes: under 25 LoC, 25-50 LoC, 50-75 LoC,
and over 75 LoC of equivalent Verilog. A clear degradation in performance is
evident as the required code size increases.


As demonstrated in Section~\ref{sec:hdlagent_eval}, \gptt is the best-performing
LLM with HDLAgent, achieving average success rates of 72\% for Chisel, 60\% for PyRTL,
64\% for DSLX, and 76\% for Verilog. Figure~\ref{fig:locoverall} reveals that for
small problems, HDLAgent performs consistently across all HDLs, with over 90\%
success rate. Conversely, for large problems exceeding 75 LoC, all HDLs,
excluding Chisel, have a consistently low 20\% success rate. The performance
difference between HDLAgent and Verilog is most pronounced in medium-sized
problems ranging from 25 to 75 LoC.

As previously mentioned, most of the errors are semantic. Interestingly, for
more complex output problems, new HDLs perform equal to Verilog even though the
LLMs have larger training in Verilog.

\begin{figure}[htbp]
\begin{center}
\includegraphics[width=0.34\textwidth]{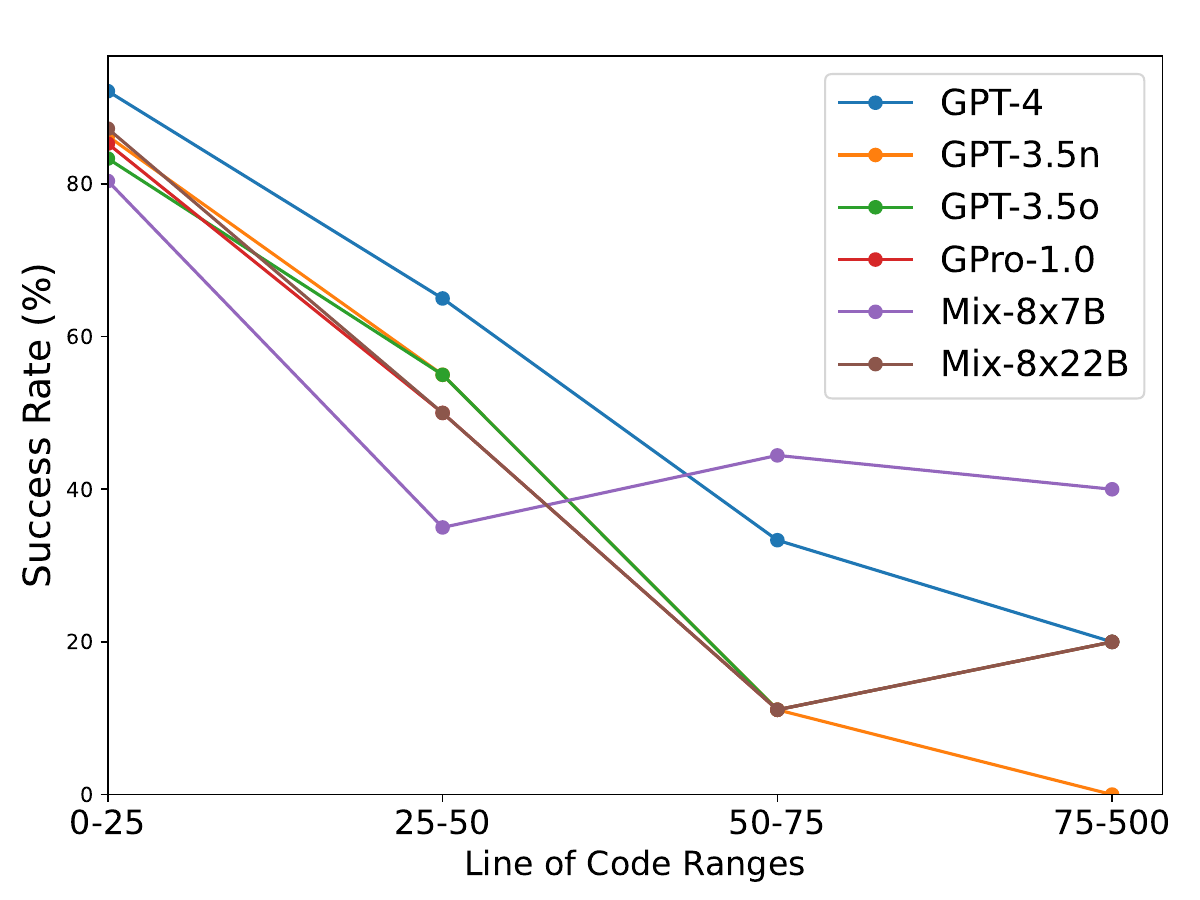}
\end{center}
\caption{HDLEval-Comb performance degrades for larger Verilog codes across LLMs.\label{fig:verilogtokens}}
\end{figure}

It is also interesting to compare across LLMs for a single HDL.
Figure~\ref{fig:verilogtokens} presents Verilog success rate with different
LLMs. While different LLMs exhibit slightly different curves, the overall trend
remains consistent: performance significantly degrades as the output problem
requires more lines of code. Another interesting observation is that Mixtral
models (\mixmoe and \mixlarge) seems to improve performance for large problems.
Although we attribute this improvement primarily to random variation, it indicates
that these models demonstrate greater resilience to large problems.

These results have two important implications: First, solving larger problems
remains a challenge that LLMs have yet to address, as evidenced by
\mixmoe achieving a success rate below 40\% even with HDLAgent enabled. Second, HDLAgent facilitates
equivalent performance for small examples across different HDLs. This latter
contribution is particularly important as it demonstrates the utility of HDLAgent for new HDL learners
querying an LLM for small code snippet generation.



\subsection{Insights for HDLs at the age of LLMs}
\label{sec:fixinghdls_eval}

The goal of this section is to show shortcomings in HDLs that must
addressed to improve accuracy in an LLM world.

\subsubsection{Verilog}

Verilog is the language that LLMs understand the best. For top-performing LLMs like
\gptt, the main challenge lies in handling pipelining. Verilog allows for unrestricted pipelining, which deviates from
the traditional Von Neumann architecture and non-hardware program structure.
\gptt effectively generates combinational logic because a typical program
without recursion or memory access can be directly translated to Verilog.
Improving pipelining remains an open research question
that must be addressed to enhance the performance of LLMs in hardware design tasks.

\subsubsection{Chisel}

Besides the common pipelining issue, Chisel LLM code generation needs help with
matching Chisel generated Verilog to native Verilog. As a part of compilation process, the generated Verilog
module's IO appends "io\_" to all names. Additional clock and reset
signals are created by default, even if unused in the original Chisel code.
Listing~\ref{lst:chisel} shows the resulting Verilog from a compiled Chisel
implementaion of a full adder circuit.

To interface modules, HDLAgent adjusts the IO to perform testing.
Postprocessing is used to remove the unused signals as well as renaming those
modified to their originals to match the circuit specification. This is
necessary as the first step of the LEC checks that the two modules' IOs match,
otherwise a truthful comparison is impossible and the LEC fails.

Additionally, both Chisel and PyRTL, being DSLs suffer from LLMs using incorrect syntax.

\begin{code}[htbp]
\begin{center}
  \begin{lstlisting}[numbers=none]
module full_adder(
  input   clock,
  input   reset,
  input   io_a,
  input   io_b,
  input   io_cin,
  output  io_sum,
  output  io_cout
);
\end{lstlisting}
\end{center}
\caption{Chisel IOs have name changes.\label{lst:chisel}}
\end{code}

\subsubsection{PyRTL}

PyRTL shares common problems with Verilog and Chisel, but it also has a problem
with semantics.

The PyRTL DSL problem is when the LLM generates Python syntax to implement logic
instead of the PyRTL syntax. In Listing~\ref{lst:pyrtl} the "INVALID" case uses Python
"inp>>1" instead of the PyRTL shift right logical library call. Many such
programs generate errors which are caught and recitifed with further HDLAgent iterations.

\begin{code}[htbp]
  \begin{center}
  \begin{lstlisting}[numbers=none]
inp = pyrtl.inpput(4, 'inp')
out = pyrtl.Output(4, 'out')
out <<= inp ^ pyrtl.shift_right_logical(inp, 1)
# equivalent: out <<= pyrtl.concat(inp[3] ^ 0, inp[3] ^ inp[2], inp[2] ^ inp[1], inp[1] ^ inp[0])
# CORRECT   : out <<= pyrtl.concat(inp[3]    , inp[3] ^ inp[2], inp[2] ^ inp[1], inp[1] ^ inp[0])
# INVALID   : out <<= inp ^ (inp>>1)  # Invalid, >> is a python shift not PyRTL
\end{lstlisting}
  \end{center}
  \caption{PyRTL issues generating right shift.\label{lst:pyrtl}}
\end{code}

Besides DSL problems, PyRTL has errors due to inconsistent semantics. In
Verilog and Chisel, a right shift logical of a positive number reduces the
bus size. For example if "inp" has 4 bits, and it is right shifted once, the output
has 3 bits. Whereas in PyRTL, it stays 4 bits but the most significant bit is hardwired to zero.
Listing~\ref{lst:pyrtl} showcases the problem in one HDLEval test. The most
significant bit is xored with zero which is not the expected result, as detailed in the "equivalent" case.

\subsubsection{DSLX}

\begin{code}[htbp]
  \begin{center}
  \begin{lstlisting}[numbers=none]
fn add_7_to_11() -> Outputs {
   //add values from 7 to 11 (exclusive)
   let base = u16:7;
   let res = for (i, accum): (u16, u16) in u16:0..u16:4 {
     accum + base + i
   }(u16:0);
   Outputs { result: res }
}
\end{lstlisting}
  \end{center}
    \caption{Rust DSLX special loop syntax.\label{lst:dslx}}
\end{code}

DSLX presented a different set of challenges than DSLs like Chisel and
PyRTL. Since it does not support unrestricted pipelining, only combinational
logic is considered in this section's feedback.

DSLX shares IO generation issues with Chisel and PyRTL but faces even
greater challenges. DSLX generated Verilog modules have a single output named "out".
DSLX's solution to multiple outputs is to return a
struct. HDLAgent addresses it by post-processing the generated Verilog and
modifying the output port name to match the desired IO. A better solution
that requires DSLX semantic changes would be to adopt a Go-like syntax that
allows for multiple named outputs and ensures Verilog generation respects those
outputs.

Another interesting source of errors stems from DSLX being "similar to Rust".
If the HDLAgent's HDL Description mentions that "DSLX is similar to Rust..." it
frequently erroneous code. Even without this sentence, the LLM sometimes generates
legal Rust but illegal DSLX code. Some differences are easy to spot, such as
DSLX's "assert(cond)" versus Rust's "assert\_eq!(cond)," while others, like the
presence of Rust annotations like "\#[test]" in DSLX code, are more subtle. To
address the "similar but not the same" syntax issues, it is suggested to avoid
mentioning the similarity and catch any discrepancies during compilation time,
generating a compile error for HDLAgent to fix.

A more complicated case involves semantic changes. Since DSLX cannot describe
circuits with mutable variables, its expressions cannot describe state changes
over a loop, making it incompatible with the Rust loop semantics. Instead,
these expressions have an accumulator value separate from the iterator,
creating a return value calculated by the body of the for loop. As shown in
Listing~\ref{lst:dslx}, the for loop body sums the values between 7 and 11 by
accumulating the base value of 7 and the iterator value in the range of 0 to
4 each loop "iteration." This deviation from standard loop semantics
required a dedicated code snippet and explanation in both the initial and
supplemental contexts to correct the LLM's often incorrect assumptions about
DSLX's generative for loop syntax. Addressing these changes will help LLMs to perform better with less HDLAgent iterations.


\section{Future Work and Conclusions}
\label{sec:conclusions}

This paper has demonstrated that Large Language Models (LLMs) hold 
transformative potential for computer science, particularly in the 
domain of Hardware Description Languages (HDLs). We introduced HDLAgent, 
an AI Agent designed to significantly enhance the ability of LLMs to 
generate code for HDLs that are not commonly represented in training datasets, 
such as Chisel, PyRTL, and DSLX. The development of new HDLs often 
relies on the capabilities of LLMs, and HDLAgent facilitates this by 
enabling effective use of existing LLMs without extensive retraining.

Our evaluations show that HDLAgent achieves a success rate of over 90\% 
on concise examples across all HDLs, making it an excellent tool for 
educational purposes in teaching new HDL languages. However, the 
performance of HDLAgent and traditional LLM approaches tends to decline 
with larger or more complex designs. For instance, even advanced LLMs 
like \gptt see a drop in success rates for Verilog projects exceeding 75 lines of code.

This work identifies several challenges and avenues for future research:
\begin{itemize}
    \item Quality of Results (QoR) issues observed in specific languages like DSLX need addressing to improve the robustness of generated designs.
    \item Consistently low success rates for pipelined designs suggest a need for specialized techniques or enhancements in LLM architectures.
    \item To accommodate complex designs, we recommend further development towards making HDLs and their compilers more conducive to LLM integration.
\end{itemize}

Moreover, while HDLAgent has shown to elevate performance significantly—raising 
the Verilog success rate of \gptt from 34\% to 72\%—it also 
highlights the scalability challenges when tackling more extensive and intricate designs.

In conclusion, HDLAgent not only broadens the applicability of LLMs 
in the field of HDLs beyond Verilog but also illuminates key 
challenges when scaling to larger systems. To aid the community and 
foster further research, we will open-source the HDLAgent code, 
providing a valuable resource for developers and researchers 
aiming to enhance the interaction between LLMs and HDL design.

\bibliographystyle{plain}
\footnotesize
\bibliography{biblio}

\begin{thebibliography}{10}

\bibitem{aianalysis}
{Artificial Analysis}.
\newblock \url{https://artificialanalysis.ai/models/gpt-35-turbo}.
\newblock Online; accessed on April 2024.

\bibitem{slang}
{slang - SystemVerilog Language Services}.
\newblock \url{https://github.com/MikePopoloski/slang}.
\newblock Online; accessed on 5 August 2021.

\bibitem{hdlbits}
{HDLBits - Verilog Practice}.
\newblock website, November 2017.

\bibitem{cwhy}
{CWhy}.
\newblock website, November 2023.

\bibitem{ahmed2022few}
Toufique Ahmed and Premkumar Devanbu.
\newblock Few-shot training llms for project-specific code-summarization.
\newblock {\em arXiv preprint arXiv:2207.04237}, 2022.

\bibitem{chisel}
Jonathan Bachrach, Huy Vo, Brian Richards, Yunsup Lee, Andrew Waterman, Rimas
  Avi{\v{z}}ienis, John Wawrzynek, and Krste Asanovi{\'c}.
\newblock Chisel: constructing hardware in a scala embedded language.
\newblock In {\em DAC Design Automation Conference 2012}, pages 1212--1221.
  IEEE, 2012.

\bibitem{chen2023teaching}
Xinyun Chen, Maxwell Lin, Nathanael Schärli, and Denny Zhou.
\newblock Teaching large language models to self-debug, 2023.

\bibitem{clow2017pyrtl}
John Clow, Georgios Tzimpragos, Deeksha Dangwal, Sammy Guo, Joseph McMahan, and
  Timothy Sherwood.
\newblock A pythonic approach for rapid hardware prototyping and
  instrumentation.
\newblock In {\em Field Programmable Logic and Applications (FPL), 2017 27th
  International Conference on}, pages 1--7. IEEE, 2017.

\bibitem{dong2023selfcollaboration}
Yihong Dong, Xue Jiang, Zhi Jin, and Ge~Li.
\newblock Self-collaboration code generation via chatgpt, 2023.

\bibitem{efabless1}
{efabless}.
\newblock Efabless 1st competition winners.
\newblock \url{https://efabless.com/genai/challenges/1}, 2023.

\bibitem{efabless2}
{efabless}.
\newblock Efabless 2nd competition winners.
\newblock \url{https://efabless.com/genai/challenges/2-winners}, 2023.

\bibitem{fan2022automated}
Zhiyu Fan, Xiang Gao, Abhik Roychoudhury, and Shin~Hwei Tan.
\newblock Improving automatically generated code from codex via automated
  program repair.
\newblock {\em arXiv preprint arXiv:2205.10583}, 2022.

\bibitem{dslx}
Google.
\newblock {XLS Website}.
\newblock {https://github.com/google/xls/}, 2022.

\bibitem{islam2023financebench}
Pranab Islam, Anand Kannappan, Douwe Kiela, Rebecca Qian, Nino Scherrer, and
  Bertie Vidgen.
\newblock Financebench: A new benchmark for financial question answering, 2023.

\bibitem{jiang2023selfplanning}
Xue Jiang, Yihong Dong, Lecheng Wang, Zheng Fang, Qiwei Shang, Ge~Li, Zhi Jin,
  and Wenpin Jiao.
\newblock Self-planning code generation with large language models, 2023.

\bibitem{kumar2024llmpowered}
Amit Kumar, Deepak Singh, Nalini Gupta, and Meena Bhatia.
\newblock Towards llm-powered verilog rtl assistant: Self-verification and
  self-correction, 2024.

\bibitem{rag}
Patrick Lewis, Ethan Perez, Aleksandra Piktus, Fabio Petroni, Vladimir
  Karpukhin, Naman Goyal, Heinrich Küttler, Mike Lewis, Wen tau Yih, Tim
  Rocktäschel, Sebastian Riedel, and Douwe Kiela.
\newblock Retrieval-augmented generation for knowledge-intensive nlp tasks,
  2021.

\bibitem{lin2023unlocking}
Bill~Yuchen Lin, Abhilasha Ravichander, Ximing Lu, Nouha Dziri, Melanie Sclar,
  Khyathi Chandu, Chandra Bhagavatula, and Yejin Choi.
\newblock The unlocking spell on base llms: Rethinking alignment via in-context
  learning, 2023.

\bibitem{liu2022few}
Haokun Liu, Derek Tam, Mohammed Muqeeth, Jay Mohta, Tenghao Huang, Mohit
  Bansal, and Colin~A Raffel.
\newblock Few-shot parameter-efficient fine-tuning is better and cheaper than
  in-context learning.
\newblock {\em Advances in Neural Information Processing Systems},
  35:1950--1965, 2022.

\bibitem{verilogeval}
Mingjie Liu, Nathaniel Pinckney, Brucek Khailany, and Haoxing Ren.
\newblock Verilogeval: Evaluating large language models for verilog code
  generation.
\newblock {\em arXiv preprint arXiv:2309.07544}, 2023.

\bibitem{liu2024rtlcoder}
Shang Liu, Wenji Fang, Yao Lu, Qijun Zhang, Hongce Zhang, and Zhiyao Xie.
\newblock Rtlcoder: Outperforming gpt-3.5 in design rtl generation with our
  open-source dataset and lightweight solution, 2024.

\bibitem{lu2023rtllm}
Yao Lu, Shang Liu, Qijun Zhang, and Zhiyao Xie.
\newblock Rtllm: An open-source benchmark for design rtl generation with large
  language model, 2023.

\bibitem{madaan2023self}
Aman Madaan, Niket Tandon, Prakhar Gupta, Skyler Hallinan, Luyu Gao, Sarah
  Wiegreffe, Uri Alon, Nouha Dziri, Shrimai Prabhumoye, Yiming Yang, Shashank
  Gupta, Bodhisattwa~Prasad Majumder, Katherine Hermann, Sean Welleck, Amir
  Yazdanbakhsh, and Peter Clark.
\newblock Self-refine: Iterative refinement with self-feedback, 2023.

\bibitem{moon2023coffee}
Seungjun Moon, Yongho Song, Hyungjoo Chae, Dongjin Kang, Taeyoon Kwon, Kai~Tzu
  iunn Ong, Seung won Hwang, and Jinyoung Yeo.
\newblock Coffee: Boost your code llms by fixing bugs with feedback, 2023.

\bibitem{ni2023lever}
Ansong Ni, Srini Iyer, Dragomir Radev, Ves Stoyanov, Wen-tau Yih, Sida~I Wang,
  and Xi~Victoria Lin.
\newblock Lever: Learning to verify language-to-code generation with execution.
\newblock {\em arXiv preprint arXiv:2302.08468}, 2023.

\bibitem{olausson2023selfrepair}
Theo~X. Olausson, Jeevana~Priya Inala, Chenglong Wang, Jianfeng Gao, and
  Armando Solar-Lezama.
\newblock Is self-repair a silver bullet for code generation?, 2023.

\bibitem{pan2010survey}
Sinno~Jialin Pan and Qiang Yang.
\newblock A survey on transfer learning.
\newblock {\em IEEE Transactions on knowledge and data engineering},
  22(10):1345--1359, 2010.

\bibitem{scholtz1989study}
Jean~Clarice Scholtz.
\newblock {\em A study of transfer of skill between programming languages}.
\newblock The University of Nebraska-Lincoln, 1989.

\bibitem{tambon2024bugs}
Florian Tambon, Arghavan~Moradi Dakhel, Amin Nikanjam, Foutse Khomh, Michel~C.
  Desmarais, and Giuliano Antoniol.
\newblock Bugs in large language models generated code: An empirical study,
  2024.

\bibitem{10137086}
Shailja Thakur, Baleegh Ahmad, Zhenxing Fan, Hammond Pearce, Benjamin Tan,
  Ramesh Karri, Brendan Dolan-Gavitt, and Siddharth Garg.
\newblock Benchmarking large language models for automated verilog rtl code
  generation.
\newblock In {\em 2023 Design, Automation and Test in Europe Conference and
  Exhibition (DATE)}, pages 1--6, 2023.

\bibitem{thakur2023verigen}
Shailja Thakur, Baleegh Ahmad, Hammond Pearce, Benjamin Tan, Brendan
  Dolan-Gavitt, Ramesh Karri, and Siddharth Garg.
\newblock Verigen: A large language model for verilog code generation, 2023.

\bibitem{thakur2023autochip}
Shailja Thakur, Jason Blocklove, Hammond Pearce, Benjamin Tan, Siddharth Garg,
  and Ramesh Karri.
\newblock Autochip: Automating hdl generation using llm feedback, 2023.

\bibitem{tsai2024rtlfixer}
Yun-Da Tsai, Mingjie Liu, and Haoxing Ren.
\newblock Rtlfixer: Automatically fixing rtl syntax errors with large language
  models, 2024.

\bibitem{wang2023intervenor}
Hanbin Wang, Zhenghao Liu, Shuo Wang, Ganqu Cui, Ning Ding, Zhiyuan Liu, and
  Ge~Yu.
\newblock Intervenor: Prompt the coding ability of large language models with
  the interactive chain of repairing, 2023.

\bibitem{wang2023chatcoder}
Zejun Wang, Jia Li, Ge~Li, and Zhi Jin.
\newblock Chatcoder: Chat-based refine requirement improves llms' code
  generation, 2023.

\bibitem{wei2023chainofthought}
Jason Wei, Xuezhi Wang, Dale Schuurmans, Maarten Bosma, Brian Ichter, Fei Xia,
  Ed~Chi, Quoc Le, and Denny Zhou.
\newblock Chain-of-thought prompting elicits reasoning in large language
  models, 2023.

\bibitem{yosys}
Clifford Wolf.
\newblock {Yosys Open SYnthesis Suite}.
\newblock \url{https://github.com/YosysHQ/yosys}, 2022.
\newblock {Online; accessed on December 2022}.

\bibitem{xia2023conversational}
Chunqiu~Steven Xia and Lingming Zhang.
\newblock Conversational automated program repair.
\newblock {\em arXiv preprint arXiv:2301.13246}, 2023.

\bibitem{yang2023large}
Chengrun Yang, Xuezhi Wang, Yifeng Lu, Hanxiao Liu, Quoc~V. Le, Denny Zhou, and
  Xinyun Chen.
\newblock Large language models as optimizers, 2023.

\bibitem{yang2023new}
Kaiyuan Yang, Haotian Liu, Yuqin Zhao, and Tiantai Deng.
\newblock A new design approach of hardware implementation through natural
  language entry.
\newblock {\em IET Collaborative Intelligent Manufacturing}, 5(4):e12087, 2023.

\bibitem{yao2023react}
Shunyu Yao, Jeffrey Zhao, Dian Yu, Nan Du, Izhak Shafran, Karthik Narasimhan,
  and Yuan Cao.
\newblock React: Synergizing reasoning and acting in language models, 2023.

\bibitem{yao2024hdldebugger}
Xufeng Yao, Haoyang Li, Tsz~Ho Chan, Wenyi Xiao, Mingxuan Yuan, Yu~Huang, Lei
  Chen, and Bei Yu.
\newblock Hdldebugger: Streamlining hdl debugging with large language models,
  2024.

\bibitem{yin2023lumos}
Da~Yin, Faeze Brahman, Abhilasha Ravichander, Khyathi Chandu, Kai-Wei Chang,
  Yejin Choi, and Bill~Yuchen Lin.
\newblock Lumos: Learning agents with unified data, modular design, and
  open-source llms, 2023.

\bibitem{hdleval}
Mark Zakharov, Farzaneh~Rabiei Kashanaki, and Jose Renau.
\newblock {{HDLE}val Benchmarking {LLM}s for Multiple {HDL}s}.
\newblock In {\em 1st IEEE International Workshop on LLM-Aided Design}, 2024.

\bibitem{zhang-etal-2023-self}
Kechi Zhang, Zhuo Li, Jia Li, Ge~Li, and Zhi Jin.
\newblock Self-edit: Fault-aware code editor for code generation.
\newblock In Anna Rogers, Jordan Boyd-Graber, and Naoaki Okazaki, editors, {\em
  Proceedings of the 61st Annual Meeting of the Association for Computational
  Linguistics (Volume 1: Long Papers)}, pages 769--787, Toronto, Canada, Jul.
  2023. Association for Computational Linguistics.

\bibitem{zhao2023expel}
Andrew Zhao, Daniel Huang, Quentin Xu, Matthieu Lin, Yong-Jin Liu, and Gao
  Huang.
\newblock Expel: Llm agents are experiential learners.
\newblock {\em arXiv preprint arXiv:2308.10144}, 2023.

\bibitem{zhao2024verilogcoder}
Jun Zhao, Min Lee, Ananya Rastogi, Huan Yu, Samuel Chen, and Li~Xiang.
\newblock Verilogcoder: Autonomous verilog coding agents with graph-based
  planning and abstract syntax tree (ast)-based waveform tracing tool, 2024.

\bibitem{zhong2024chatgpt}
Li~Zhong and Zilong Wang.
\newblock Can chatgpt replace stackoverflow? a study on robustness and
  reliability of large language model code generation, 2024.

\bibitem{zhou2023agents}
Wangchunshu Zhou, Yuchen~Eleanor Jiang, Long Li, Jialong Wu, Tiannan Wang, Shi
  Qiu, Jintian Zhang, Jing Chen, Ruipu Wu, Shuai Wang, Shiding Zhu, Jiyu Chen,
  Wentao Zhang, Ningyu Zhang, Huajun Chen, Peng Cui, and Mrinmaya Sachan.
\newblock Agents: An open-source framework for autonomous language agents,
  2023.

\end{thebibliography}
\end{document}